\documentclass[a4paper,11pt]{article}
\usepackage{amsmath}
\DeclareMathOperator{\arccot}{arccot}
\usepackage{jheppub}
\usepackage{amssymb}
\usepackage{graphicx}
\usepackage[utf8]{inputenc}
\usepackage{xspace}
\usepackage{color}
\usepackage{ulem}
\usepackage{array}
\usepackage{soul}

\title{Thermal Relic of Self-Interacting Dark Matter with Retarded Decay of Mediator}

\date{\today}

\author[1,2]{Bin Zhu,}

\author[3]{Murat Abdughani}

\affiliation[1]{Department of Physics, Yantai University, Yantai 264005, P. R. China}

\affiliation[2]{Department of Physics, Chung-Ang University, Seoul 06974, Korea}

\affiliation[3]{Key Laboratory of Dark Matter and Space Astronomy, Purple Mountain Observatory, Chinese Academy of Sciences, Nanjing 210023, China}

\emailAdd{zhubin@mail.nankai.edu.cn, mulati@pmo.ac.cn}

\abstract{
The existence of a light mediator is beneficial to some phenomena in astroparticle physics, such as the core-cusp problem and diversity problem. It can decouple from Standard Model to avoid direct detection constraints, generally realized by retard decay of the mediator. Their out-of-equilibrium decay process changes the dark matter (DM) freeze-out via temperature discrepancy. This type of hidden sector (HS) typically requires a precision calculation of the freeze-out process considering HS temperature evolution and the thermal average of the cross-section. If the mediator is light sufficiently, we can not ignore the s-wave radiative bound state formation process from the perspective of CMB ionization and Sommerfeld enhancement. We put large mass splitting between DM and mediator, different temperature evolution on the same theoretical footing, discussing the implication for DM relic density in this HS. We study this model and illustrate its property by considering the general Higgs-portal dark matter scenario, which includes all the relevant constraints and signals. It shows that the combination of BBN and CMB constraint favors the not-too-hot HS, $r_{\mathrm{inf}}<10^2$, for the positive cubic interaction of mediator scenario. On the other hand, the negative cubic interaction is ruled out except for our proposed blind spot scenario. 
}

\begin{document}

\maketitle


\section{Introduction}

Most of the dark matter (DM) community have considered the situation in which DM mass is comparable with mediator mass~\cite{Bertone:2004pz}, e.g. $100$ GeV neutralino with Higgs or W/Z bosons being the mediators. As a result, these processes belong to contact short-range interaction~\cite{Bertone:2004pz,Feng:2010gw}. However, with the development of indirect detection and DM N-body simulation, long-range interaction attracts lots of attention such as Positron Excess~\cite{ArkaniHamed:2008qn}. Besides, although cold dark matter (CDM) in $\Lambda$CDM model is very successful at explaining the current observations on large scales~\cite{Bahcall:1999xn}, there are various discrepancies between N-body simulations of collisionless CDM and astrophysical observations on galactic scales. Self-interacting dark matter (SIDM) was thus proposed to solve the  core-cusp~\cite{Moore:1999gc, Moore:1994yx} and missing satellites~\cite{Moore_1999, Klypin199982, Zavala20091779, 10.1111/j.1365-2966.2009.16188.x, Trujillo_Gomez_2011} problems on galactic and smaller scales. The energy transfer from self-interaction in the central regions of DM halos provides a method to heat DM particles so that an isothermal core can be created easily. In SIDM~\cite{Randall:2007ph,Buckley:2009in,Tulin:2013teo,Kaplinghat:2013yxa,Boddy:2014yra,DelNobile:2015uua,Bernal:2015ova,Tulin:2017ara}, the cross section within long-range interaction is velocity-dependent in order to escape bullet-cluster constraint at relatively high velocity and obtains large cross section at low velocity. Furthermore, Sommerfeld enhancement (SE) on the annihilation cross sections can significantly deplete DM relic density. This is mainly because DM pair annihilation occurs at small non-relativistic velocity at freeze-out temperature.

The non-observational evidence of WIMP in both direct detection and indirect detection challenges the traditional form of SIDM, which leads to the study of viable generalizations of SIDM to overcome stringent limitations such as pseudo-Dirac SIDM~\cite{Zhang:2016dck}. The basic strategy for model building is to find a mechanism that realizes correct DM abundance in the early universe without causing a large elastic scattering rate with nucleon. Hidden sector (HS) DM~\cite{Berlin:2016vnh,Tenkanen:2016jic,Okawa:2016wrr,Berlin:2016gtr,Escudero:2017yia} is one of the most attractive options following the strategy, where DM freezes out of thermal equilibrium entirely within its HS.

In this paper, for simplicity, the decoupled HS only contains two particles: DM and mediator. The large mass hierarchy between DM and mediator offers long-range interaction. As a result, SIDM can be naturally realized in this class of models. In addition, radiative bound state formation (BSF) and SE are inevitable, which lead to large annihilation cross section. 
To avoid large elastic scattering between DM and nucleon, we hypothesize the coupling between the mediator and visible sector (VS) particles is ultra-tiny so that the HS is completely decoupled from the VS thermal bath. As a consequence, HS have its own temperature, which may also increase the annihilation cross section. 
The mediator provides a portal between the HS and VS, and can decay into VS particles and connect the two sectors. 
Unlike other models, we can achieve correct velocity dependence at late times without assuming particle-antiparticle asymmetry in the HS and can get the full relic abundance. 
\textit{The purpose of this paper is thus to perform a precise calculation on the HS relic density with large mass splitting by illustrating General Higgs Portal Dark Matter (GHPDM) as our benchmark model of the HS.}

The structure of this paper is organized as follows: In Sec.~\ref{sec:model} We derive the temperature ratio between the HS and the VS in terms of entropy conservation, and implement it to the Boltzmann equation. And then briefly review the general contents of our benchmark model, GHPDM, in which SE and formation of bound state can be realized. The constraints from BBN an CMB are also calculated.
We give the semi-analytical formula to calculate the self-interaction cross-section in Sec.~\ref{sec:sidm}. A simple chi-squared analysis shows that mass ratio $m_{\phi}/m_{m_{\chi}}$ smaller than $10^{-4}$ is favored.  In Sec.~\ref{sec:cross}, we show the analytical results of SE and radiative BSF cross-section. The implication of relevant DM properties such as relic abundance involving SE effect and BSF are shown explicitly. We finally conclude in Sec.~\ref{sec:conclusion}.

\section{Boltzmann equation in the hidden sector}
\label{sec:model}
\subsection{Hidden sector temperature and Boltzmann equation}
When DM and mediator as a system  effectively decoupled from SM thermal bath, the two sectors undergo different temperature evolution, which is convenient to define the ratio of the HS and VS temperatures, $r=T_h/T$. Here quantities with subscript or superscript $h$ stand for HS. We assume both sectors are populated after inflation (reheating). Therefore the ratio of their initial temperature $r_{\mathrm{inf}}=T_{\mathrm{inf}}^h/T_{\mathrm{inf}}$ can be regarded as a free parameter and initial condition. With the expansion of the universe, the evolution of temperature is determined by entropy densities namely $s=\left(2 \pi^{2} / 45\right) g_{*s}(T) T^{3}$ and $s_{h}=\left(2 \pi^{2} / 45\right) g_{*s}^{h}\left(T_{h}\right) T_{h}^{3}$, where $g_{*s} (T)$ and $g_{*s}^h (T_h)$ are the effective relativistic entropy degrees-of-freedoms (d.o.f.) of VS and HS. Even though the temperatures of the two sectors are independent of each other, the conservation of comoving entropy densities, $d(sa^3)/dt=0$, determines the evolution of ratio as follows\cite{Berlin:2016gtr,Bringmann:2020mgx}
\begin{equation}
r=\left(\frac{g_{*s}(T)}{g_{*s, \inf} }\right)^{1 / 3}\left(\frac{g^{h}_{*s, \inf}}{g_{*s}^{h}\left(T_{h}\right)}\right)^{1 / 3} r_{\inf} \label{eq:r}
\end{equation}
where $g_{*s,\inf}$ and $g^h_{*s,\inf}$ are relativistic entropy d.o.f. at inflation for VS and HS. To solve for the temperature ratio $r$ in Eq.~(\ref{eq:r}), we need to present the evolution of the d.o.f. $g_{*}$ and set up the initial condition $r_{\mathrm{inf}}$. Different initial conditions lead to different DM phenomenology which will be discussed in Sec.~\ref{sec:cross}. 

Assume that the temperature at inflation is much higher than the mass of any particle in two sectors, then d.o.f. at inflation equal $\sum (g_s + \frac{7}{8} g_f)$, where $g_s$ and $g_f$ are the intrinsic d.o.f. of scalars and fermions. For example, if there are one complex scalar $\phi$ and one Majorana fermion $\chi$ in HS, then $g_{*s,\inf}^h = 3.75$. If VS only consists of SM particles, then $g_{*s,\inf} \simeq 106.75$. Evolution of $g_{*s}(T)$ for SM particles is given in~\cite{Husdal:2016haj}. Assuming the chemical potential are negligible, effective entropy d.o.f. $g_{*s}$ as a function of T for a particle with a mass $m$ and intrinsic d.o.f. $g$ can be obtained by
\begin{equation}
    \begin{aligned}
    g_{*\epsilon} &= \frac{15g}{4\pi^4} \int_{z}^\infty \frac{u^2 \sqrt{u^2-z^2}}{e^u \pm 1} du ~,
    \\
    g_{*p} &= \frac{15g}{4\pi^4} \int_{z}^\infty \frac{(u^2-z^2)^{3/2}}{e^u \pm 1} du ~,
    \\
   g_{*s} &= \frac{3g_{*\epsilon}+g_{*p}}{4} ~,
    \end{aligned}
\end{equation}
where $z = m/T $ and positive sign in the formula stands for fermion and negative for scalar particle.

\begin{figure}[!htbp]
	\centering
	\includegraphics[width=0.65\textwidth]{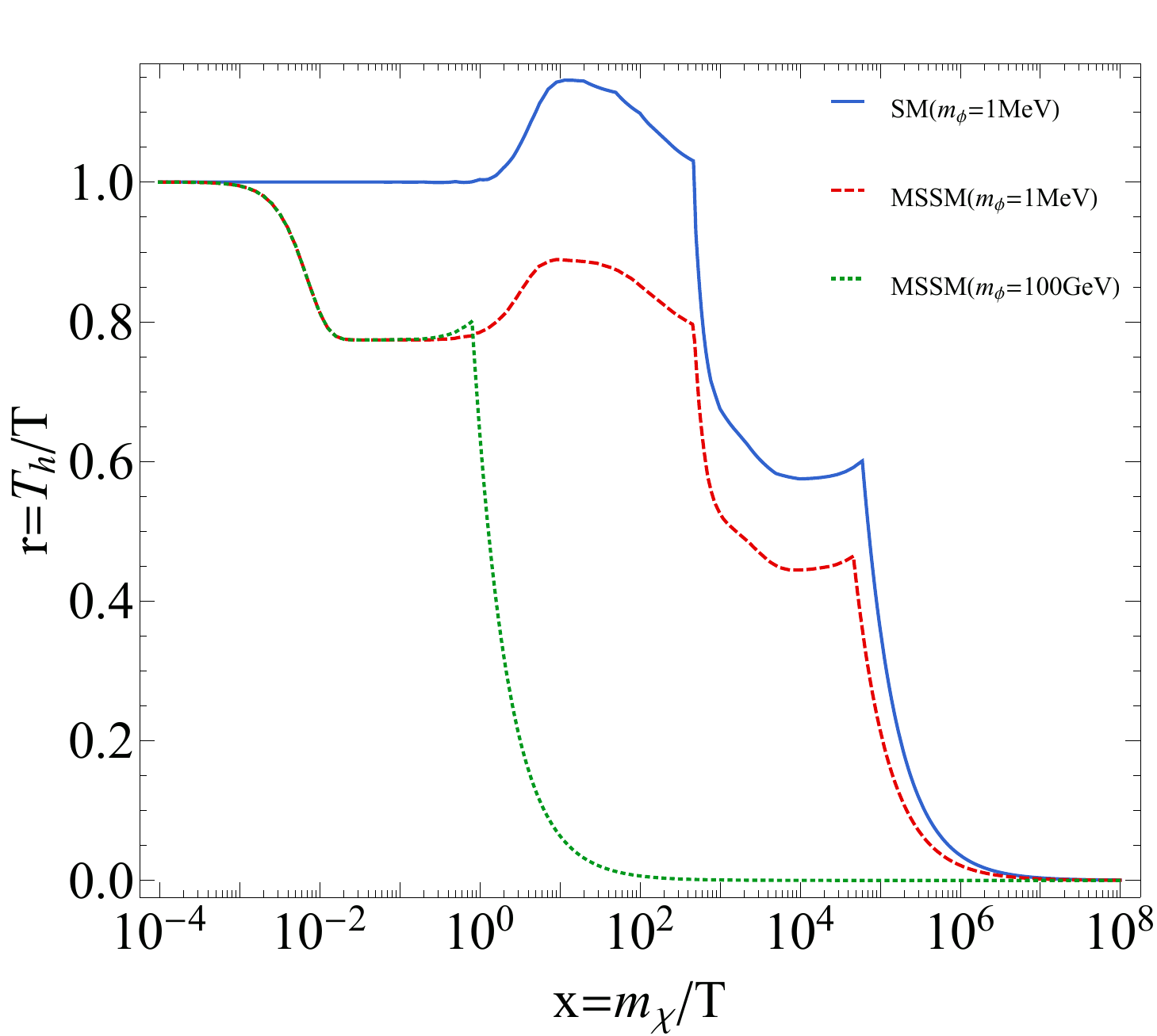}
    \caption{Evolution of HS and VS temperature ratio $r$ as a function of $x=m_\chi/T$ for different choice of parameters. Blue solid line: The VS only consists of SM particles. Red dashed line: The VS consists of MSSM particles, all MSSM particles with the mass of 50 TeV. Green dotted line: Same as the red line, but the mediator mass $m_\phi$ equals to $m_\chi = 100$ GeV.
	}
	\label{fig:r}
	\end{figure}

The Eq.~(\ref{eq:r}) can be solved after determining the variation of $g_{*s}(T)$ and $g_{*s}^h(T_h)$. In Fig.~\ref{fig:r}, we show the evolution of the $r$ with respective to the $x=m_\chi/T$ for the different choice of parameters. Here, as an example, we adopt minimal supersymmetric standard model (MSSM) with all supersymmetric particle masses of 50 TeV in the VS, and its d.o.f at inflation is $g_{*s,\inf} \simeq 228.75$. The first drop of $r$ values around $x = 10^{-3}$ are the decoupling of supersymmetric particles. The ratio $r$ in blue and orange line start to increase after $x=1$, since DM becomes non-relativistic. In green line, $m_\phi$ as heavy as DM mass $m_\chi$ = 100 GeV, they 
became non-relativistic simultaneously. After the decoupling of lightest particles in HS, which is the mediator $\phi$ in our case, the comoving number density is conserved, equally phase space density $f \sim e^{-E/T_h}$ is constant. From $E \sim a^{-2}$ and $T \sim a^{-1}$, we know that
\begin{equation}
   r \sim T \sim a^{-1}~,
\end{equation}
and it can be seen from the tail of the curves in Fig.~\ref{fig:r}.

Within the temperature ratio evolution $r$ known, the Boltzmann equation describing HS $\chi,\phi$ before entropy injection from mediator is straightforward to obtain

\begin{equation}
\begin{aligned}
\dot{n}_{\chi}+3 H n_{\chi}&=-\langle\sigma v\rangle_{\chi\chi \rightarrow \phi\phi}\left[n_{\chi}^{2}-n_{\phi}^{2}\left(\frac{n_{\chi}^{\mathrm{eq}}}{n_{\phi}^{\mathrm{eq}}}\right)^{2}\right]\\
\dot{n}_{\phi}+3 H n_{\phi}&=\langle\sigma v\rangle_{\chi\chi \rightarrow \phi\phi}\left[n_{\chi}^{2}-n_{\phi}^{2}\left(\frac{n_{\chi}^{\mathrm{eq}}}{n_{\phi}^{\mathrm{eq}}}\right)^{2}\right]-\Gamma_{\phi}\left[n_{\phi}-n_{\phi}^{\mathrm{eq}}\right]
\end{aligned}
\label{eqn:Boltzmann}
\end{equation}

If $\chi$ is not self-conjugated, we will replace $\langle\sigma v\rangle$ by $\langle\sigma v\rangle/2$. We make some general comments on the Boltzmann equation: 
\begin{itemize}
\item When the DM mass is almost degenerate with mediator mass, Boltzmann equations in~\eqref{eqn:Boltzmann} reduce to Co-decaying DM~\cite{Dror:2016rxc}. However, Self-interacting DM requires the mediator $\phi$ much lighter than DM $\chi$. The large hierarchy is the main difference between our model and Co-decaying DM. Due to the existence of long-range interaction, the cross-section $\langle\sigma v\rangle_{\chi\chi\rightarrow \phi\phi}$ includes both SE and BSF effects. 

\item When the decay of $\phi$ into SM particles is prompt, the HS remains thermal equilibrium with the VS. In that sense, only one temperature in the Boltzmann equation $T_h=T$ appears.  However, we assume the mediator decay is too tiny to have direct detection constraints. The HS and VS undergo different temperature evolution described by the ratio $r$. For the varying values of $r$, the corresponding solution becomes different. Since the portal coupling between mediator $\phi$ and SM sector is ultra-tiny, mediator decay is retarded decay process. The out-of-equilibrium decay injects more entropy to the thermal bath and violates entropy conservation. Therefore our equation is only valid up to the time that the mediator decay process dominates. Furthermore, additional entropy injection with the early matter-dominated universe will dilute the pre-existing relic density of DM. In our setup, the dilution factor $\Delta$ is not sizeable because of the small mediator mass. 

\begin{equation}
    \Delta=\frac{S_{f}}{S_{i}}=\frac{T_{f}^{3}}{T_{i}^{3}} \approx 2.1 g_{*}^{1 / 4} \frac{m_{\phi} Y_{\phi} \tau_{\phi}}{M_P}^{1/2}
\end{equation}

\item Since DM and mediator annihilate with each other efficiently to maintain the kinetic equilibrium, the number density of mediator $n_{\phi}$ tracks with thermal equilibrium $n_{\phi,\mathrm{eq}}$. We will calculate the elastic scattering process $\chi\phi\rightarrow\chi\phi$ in Sec.~\ref{sec:cross} to prove it. Then, the Boltzmann equation~\eqref{eqn:Boltzmann} reduces to conventional WIMP type equation

\begin{equation}
    \dot{n}_{\chi}+3 H n_{\chi}=-\langle\sigma v\rangle_{\chi\chi \rightarrow \phi\phi}(n_{\chi}^{2}-n_{\chi,\mathrm{eq}}^{2})
    \label{eqn:WIMP}
\end{equation}

There are some modifications that we need to consider in Eq.~\eqref{eqn:WIMP}. One is that we should perform the thermal average of cross section $\langle\sigma v\rangle$ in $T_h$ rather than T. Furthermore, thermal equilibrium is also modified to be evaluated at hidden temperature $T_h$,

\begin{equation}
n_{\chi, \mathrm{eq}}=g_{\chi} \frac{\left(m_{\chi}^{2} T_{h}\right)}{2 \pi^{2}} K_{2}\left[\frac{m_{\chi}}{T_{h}}\right]=g_{\chi} \frac{\left(m_{\chi}^{2} r T\right)}{2 \pi^{2}} K_{2}\left[\frac{m_{\chi}}{r T}\right]\label{eq:eq}
\end{equation}
where $K_2$ is the Bessel function of the second kind. Despite conventional WIMP scenario, the Hubble constant includes the contribution of the HS,

\begin{equation}
H^{2}=\frac{8 \pi}{3 m_{\mathrm{pl}}^{2}}\left(\rho+\rho_{h}\right)=\frac{8 \pi}{3 m_{\mathrm{pl}}^{2}} \frac{\pi^{2}}{30}\left(g_{*\epsilon}(T) T^{4}+g_{*\epsilon}^{h}(T_h) T_{h}^{4}\right) \equiv \frac{4 \pi^{3}}{45} \frac{m_{\chi}^{4}}{m_{\mathrm{pl}}^{2}} \frac{g_{*}^{\mathrm{eff}}(T)}{x^{4}}\label{eq:H}
\end{equation}
where $m_\mathrm{pl} = 1.22 \times 10^{19} $ GeV is the plank mass, $g_{*\epsilon}$ and $g_{*\epsilon}^h$ are the energy d.o.f of the VS and HS respectively. 

\end{itemize}

\subsection{Benchmark model}

From here, for the specific calculation of some observables, we implement GHPDM as the benchmark model. When the HS contains only a Majorana fermion DM $\chi$ and a complex scalar mediator $s$, the relevant Lagrangian is written as follows 
\begin{equation}
-\mathcal{L}=m_s^2|s|^2+\lambda |s|^2(|h_u|^2+h_d^2)+\kappa s\chi \chi+A_{\lambda}s h_u h_d+\frac{A_{\kappa}}{3}s^3+h.c.
\end{equation}
where $h_u$ and $h_d$ stand for the up and down-type Higgs doublets respectively. The Higgs portal of the Standard Model provides the opportunity for coupling to a very light
scalar field $s$ via the renormalizable operator $|s|^2(h_u^2+h_d^2)$ and super-renormalizable operator $sh_u h_d$. It allows for the existence of large direct detection cross section and Higgs invisible decay width. To avoid these two constraints, we assume the portal coupling $\lambda$ and $A_{\lambda}$ to be ultra-tiny. The $\kappa$ quantifies the interaction between DM and mediator, while $A_{\kappa}$ describes the self-coupling between singlets. The GHPDM can be viewed as the low energy effective theory of general NMSSM. The UV complete interaction is determined by the superpotential and soft susy breaking term: 
\begin{equation}
W=(\mu+\epsilon S)H_u H_d+\frac{1}{2}\mu_s^{2}S^2+\frac{\kappa}{3}S^3
\end{equation}
and
\begin{equation}
-\mathcal{L}_{\text{soft}}=m_{h_{u}}^{2}\left|h_{u}\right|^{2}+m_{h_{d}}^{2}\left|h_{d}\right|^{2}+m_{s}^{2}|s|^{2}+\left(B \mu h_{u} h_{d}+A_{\lambda} s h_{u} h_{d}+\frac{B \mu_{s}}{2} s^{2}+\frac{A_{\kappa}}{3}  s^{3}+\mathrm{h.c.}\right)
\end{equation} 
Since we are only interested in the DM phenomenology, there is no need to consider the UV completion-supersymmetry in detail. The complex singlet mediator can be further divided into CP-even and odd part,
\begin{align}
s=\frac{1}{\sqrt{2}}(v_s+\phi+i a)
\end{align}
Therefore our HS is described in the following lagrangian.
\begin{equation}
\begin{aligned} \mathcal{L}_{f} &=\overline{\chi} i \partial^{\mu}\gamma_{\mu} \chi+\frac{1}{2} \partial_{\mu} \phi \partial^{\mu} \phi+\frac{1}{2} \partial_{\mu} a \partial^{\mu} a-m_{\chi} \overline{\chi} \chi-\frac{1}{2} m_{\phi}^{2} \phi^{2}-\frac{1}{2}m_{\phi}^2 a^2 \\ &-\frac{\kappa}{3} \phi \overline{\chi} \chi-\frac{A_{\kappa}}{3} \phi^{3}-\epsilon \phi H_u H_d
\end{aligned}
\end{equation}
The input parameters of this model are thus $\{m_{\chi}, m_{\phi}, \alpha=\kappa^2/4\pi, \epsilon,A_{\kappa}\}$. We should mention that this HS is not limited to supersymmetry but UV completion of the Higgs-portal model. The CP-odd singlet $a$ could provide a long-range interaction. One main reason for giving up this option is that the long-range potential mediated by the CP-odd singlet is the function of the spin structure which becomes negligible after averaging spin configuration~\cite{ArkaniHamed:2008qn}. As a result, we only focus on Yukawa potential mediated by CP-even singlet $\phi$,
\begin{align}
V=-\frac{\alpha}{r}\exp(-m_{\phi}r) ~.
\label{eq:yukawa}
\end{align}

\subsection{Constraints from BBN and $\Delta N_{\rm eff}$}

To avoid overclosing the universe, the mediator $\phi$ and $a$ ultimately decay into Standard Model particles. The most important constraint for its lifetime comes from Big Bang Nucleosynthesis (BBN) $\Gamma_{\phi}>10^{-25}\mathrm{GeV}$. The decay width of the mediator comes from mass-insertion calculation~\cite{Dror:2016rxc}. The allowed decay channels are almost the same as Higgs decay if the kinematics channel is open such as $\phi\rightarrow f\bar{f}, gg,\gamma\gamma$. Here we only list $\phi\rightarrow gg$ for brevity. 

\begin{equation}
\Gamma_{g g}^{\phi}=\frac{\epsilon^{2} v^{2} m_{h}\left(2 \frac{\mu}{m_{\phi}}-s_{2 \beta}\right)^{2}}{2m_{\phi}^{3}\left(1-\frac{m_{h}^{2}}{m_{\phi}^{2}}\right)^{2}} \Gamma_{g g}^{h}
\end{equation}
	
\begin{figure}[!htbp]
	\centering
	\includegraphics[width=0.65\textwidth]{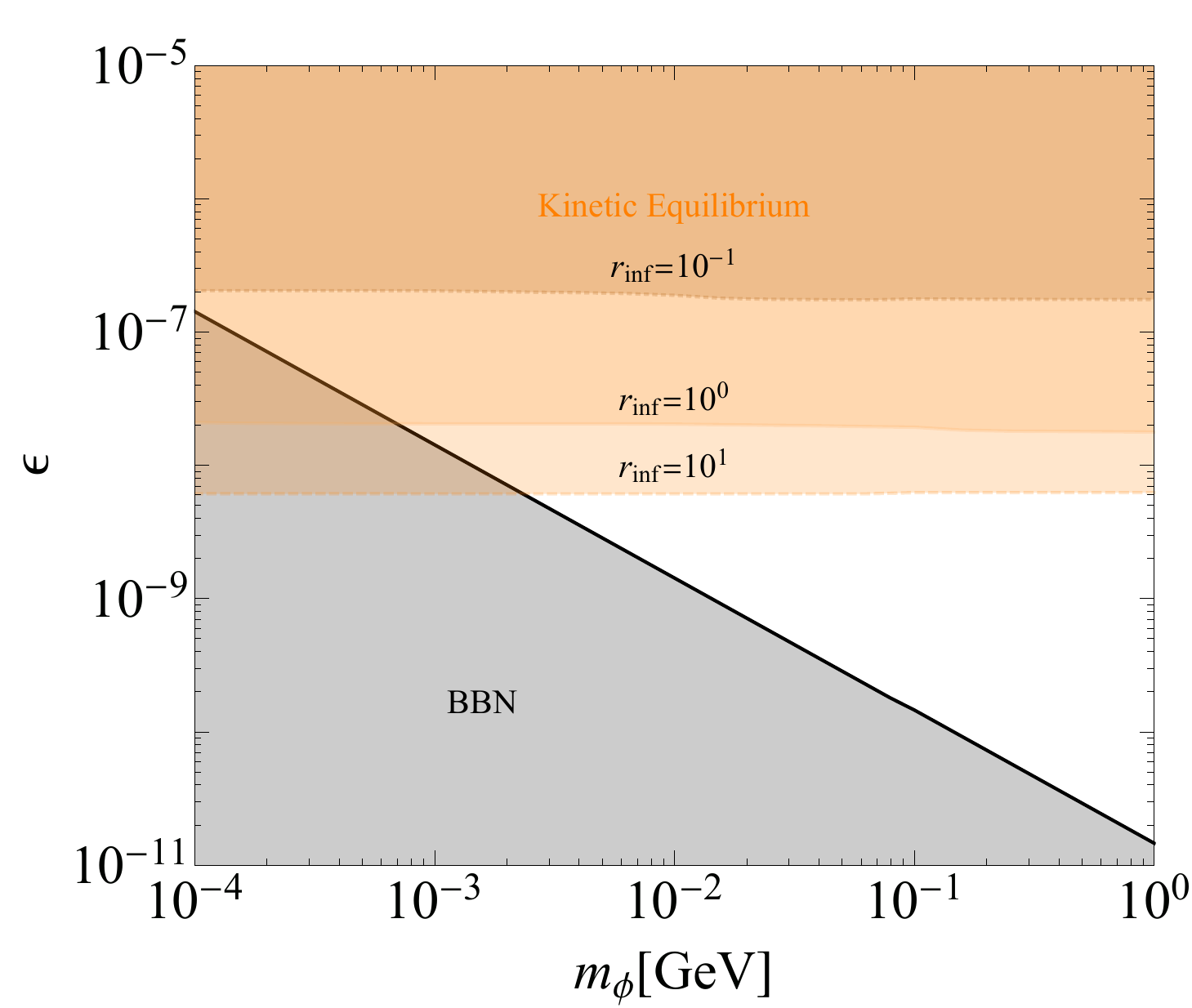}
	\caption{Constraints on the $\phi$ versus $\epsilon$ parameter space. The gray shaded region indicates where decays occur after neutrino decoupling and is excluded by BBN. The shaded region above (below) the orange dashed curve indicates that the decays of $\phi$ occur in (out) of equilibrium.
	}
	\label{fig:Fig2}
	\end{figure}

Constraints from BBN are shown in Fig.~\ref{fig:Fig2}. The grayed shaded region corresponds to the decays occurring after BBN. An additional requirement of HS places upper bound on the coupling $\epsilon$. The orange shaded regions are excluded assuming that the HS remains thermal equilibrium with SM thermal bath via the condition $\Gamma\sim 0.1 H_{\mathrm{eff}}$ $(T_h=m_{\phi})$. 

The energy density of the mediator affects the measurement of the effective number of relativistic neutrino species $N_{\mathrm{eff}}$. Current CMB experiment thus can probe our HS via $\Delta N_{\mathrm{eff}}<0.26$. After neutrino decoupling, $\Delta N_{\mathrm{eff}}$ can be obtained without considering decay product, 

\begin{equation}
\Delta N_{\mathrm{eff}} =\frac{4}{7}\left(\frac{11}{4}\right)^{4 / 3} g_{*}^{h}(T_h) r_{h}^{4} 
\end{equation}

It is easy to find that the maximal $\Delta N_{\mathrm{eff}}$ for our model is $0.1$ at the limit of a massless mediator. So we can safely ignore $\Delta N_{\mathrm{eff}}$ constraint. Note that if mediator dominantly decays into a neutrino, the constraint re-appears. However, the branching ratio into neutrino is tiny in the Higgs portal.

\section{Self-Interacting Dark Matter Realization}
\label{sec:sidm}

There are three main approaches to realize velocity-dependent self-interacting cross-sections. One is the existence of a light mediator to produce velocity-dependent cross-section by non-perturbative resummation. The second approach borrows the similarity of neutron-proton scattering where confinement in the HS adapts finite-size effect~\cite{Chu:2018faw} to decrease the cross-section at low velocity. Finally, resonant SIDM~\cite{Chu:2018fzy} can naturally host a velocity-dependent behavior. In our paper, we focus on the first approach.

When the momentum of DM $k$ is much less than the mass of mediator $m_\phi$, the cross-section is mainly from the $S$-wave. However, for $k \gtrsim m_\phi$, higher angular momentum $\ell > 0$ states become more important when it goes to classical limit $k/m_\phi \to \infty$. Commonly used momentum transfer cross section~\cite{PhysRevD.81.083522,PhysRevLett.104.151301,PhysRevD.66.063002}

\begin{equation}
    \sigma_{\rm T} \equiv \int d\Omega \, (1-\cos \theta)\frac{d\sigma}{d\Omega}
\end{equation}
is more suitable than total cross-section $\sigma = \int d\Omega (d\sigma /d\Omega)$ due to the scattering velocity and angle dependence of $d\sigma /d\Omega$. However, problems will emerge when the DM is the identical particle. Therefore, viscosity cross section~\cite{Tulin:2013teo}

\begin{equation}
    \sigma_{\rm V} \equiv \int d\Omega \, \sin^2 \theta \frac{d\sigma}{d\Omega}
\end{equation}
is adopted from heat conductivity~\cite{kruger1965introduction}. Viscosity cross-section may better characterize DM halo dynamics since DM scattering is described by heat conductivity in fluid formulations of self-interacting DM~\cite{Gnedin:2000ea, Balberg:2002ue}. 

In the partial wave analysis, the differential cross section
\begin{equation}
    \frac{d\sigma}{d\Omega} = \frac{1}{k^2} \left| \sum_{\ell = 0}^{\infty} (2\ell + 1) e^{i \delta_\ell} P_\ell (\cos \theta) \sin \delta_\ell \right|^2
\end{equation}
is achieved by determining the phase shift $\delta_\ell$ via solving the Schrodinger equation
\begin{equation}
    \frac{1}{r^2} \frac{\partial}{\partial r} \left( r^2 \frac{\partial \mathcal{R}_\ell}{\partial r} \right) + \left( E - V(r) - \frac{\ell(\ell+1)}{2\mu r^2}  \right) \mathcal{R}_\ell(r) = 0 \, .
\end{equation}
Thus, the approximated viscosity cross section is obtained
\begin{equation}
    \sigma_{\rm V} m_\phi^2 \approx \frac{2\pi}{\kappa^2} \int_{1}^\infty \mathrm{d}\ell \, \ell \, \sin^2 2\delta'(\ell - 1/2)
\end{equation}

For identical particles, spatial part of the wave equation requires to be even (odd) for symmetric (anti-symmetric) spin parts. Then the differential cross section is given by
\begin{equation}
    \frac{\mathrm{d} \sigma}{\mathrm{d}\Omega} = \frac{1}{k^2} \left\lvert \sum_{\ell=0}^\infty(2\ell+1) e^{i \delta_\ell} \left[P_\ell(\cos \theta) \pm P_\ell(-\cos \theta) \right] \sin \delta_\ell \right\rvert^2 \; , \label{eq:identical}
\end{equation}
where the positive (negative) sign correspond to an even (odd) spatial wave function, respectively. From Eq.~\eqref{eq:identical} and $P_\ell (-x) = (-1)^\ell P_\ell (x)$, we know that, for even (odd) case contribution from odd (even) phase shifts vanishes. The viscosity cross sections for even and odd partial wave cases are 
\begin{equation}
 \sigma_\mathrm{V}^\text{even(odd)} \, m_\phi^2 \approx \frac{2\pi}{\kappa^2} \int_{1/2(3/2)}^\infty\mathrm{d}\ell \, \ell \sin^2 2\delta'(\ell-1/2)\, .
\end{equation}
For realistic unpolarized scattering particles, contribution to the total cross section comes both from even and odd viscosity cross section. For example, cross section for self-interacting Majorana DM is
\begin{equation}
    \sigma_\mathrm{V} = \frac{1}{4} \sigma_\mathrm{V}^\mathrm{even} + \frac{3}{4} \sigma_\mathrm{V}^\mathrm{odd} \, .
\end{equation}

The last step to get useful cross section is to find analytical solution of the phase shift $\delta_\ell$ represented in terms of 
\begin{equation}
\eta = \frac{m_\chi v}{m_\phi} \, , \quad 
\beta = \frac{2\alpha m_\phi}{m_\chi v^2} \, ,
\end{equation}
which correspond to the dimensionless momentum and strength of the potential relative to the kinetic energy, respectively. Viscosity cross section for attractive potential is~\cite{Colquhoun:2020adl}
\begin{align}
\sigma_\mathrm{V}^{\mathrm{Semi-Classical}} & = \frac{\pi}{m_\phi^2}  \times \begin{cases}
4\beta^2 \zeta_n\left(\eta, 2\beta\right) & \beta\leq0.1\\
\hspace{7cm}\  & \ \\[-4mm]
4\beta^2 \zeta_n\left(\eta, 2\beta\right) e^{0.67(\beta - 0.1)} & 0.1 < \beta \leq0.5\\
\hspace{7cm}\  & \ \\[-4mm]
2.5 \log(\beta + 1.05) & 0.5 < \beta < 25 \\
\hspace{7cm}\  & \ \\[-4mm]
\frac{1}{2} \left(1 + \log{\beta} - \frac{1}{2\log\beta}\right)^2 & \beta\geq25
\end{cases}
\end{align}
with
\begin{align}
 \zeta_n(\eta,\beta) & =\frac{\text{max}(n,\beta\eta)^2 - n^2}{2 \eta^2 \beta^2} + \eta\left(\frac{\text{max}(n,\beta\eta)}{\eta}\right)\,,\\
 \eta(x) & = x^2 \left[ -K_1\left(x\right)^2 + K_0\left(x\right) K_2\left(x\right)\right]\,,
\end{align}
where $K_0$, $K_1$ and $K_2$ are the Bessel functions of the second kind. These formula is valid in semi-classical limit i.e. $\eta>1$. When $\eta$ is smaller than $0.4$, it belongs to quantum regime where Hulthen potential can be used to solve Schrodinger equation

\begin{equation}
    \sigma_\mathrm{V}^{\mathrm{Hulthen}}=\frac{4\pi}{\eta^2}\frac{\sin^2\delta}{\pi},
\end{equation}
where the phase $\delta$ is determined as follows

\begin{equation}
\begin{aligned}
    \delta&=\arg\left(i \frac{\Gamma(l_{+} + l_{-} -2)}{\Gamma(l_{+})\Gamma(l_{-})}\right),\\
    l_{+}&=1+\frac{\eta}{1.6}(i+i\sqrt{3.2\beta-1}),\\
    l_{-}&=1-\frac{\eta}{1.6}(i+i\sqrt{3.2\beta-1}).
    \end{aligned}
\end{equation}

Within the regime $0.4<\eta<1$, we adapt the interpolation method $\sigma_{V}^{\mathrm{Interpolation}}=(1-\eta)/0.6 \sigma_{V}^{\mathrm{Hulthen}}+(\eta-0.4)/0.6 \sigma_{V}^{\mathrm{Semi-Classical}}$ in the overlap region. In terms of these three cross-sections, we can cover most of interesting parameter space of SIDM. 

\begin{figure}[!htbp]
	\centering
	\includegraphics[width=0.65\textwidth]{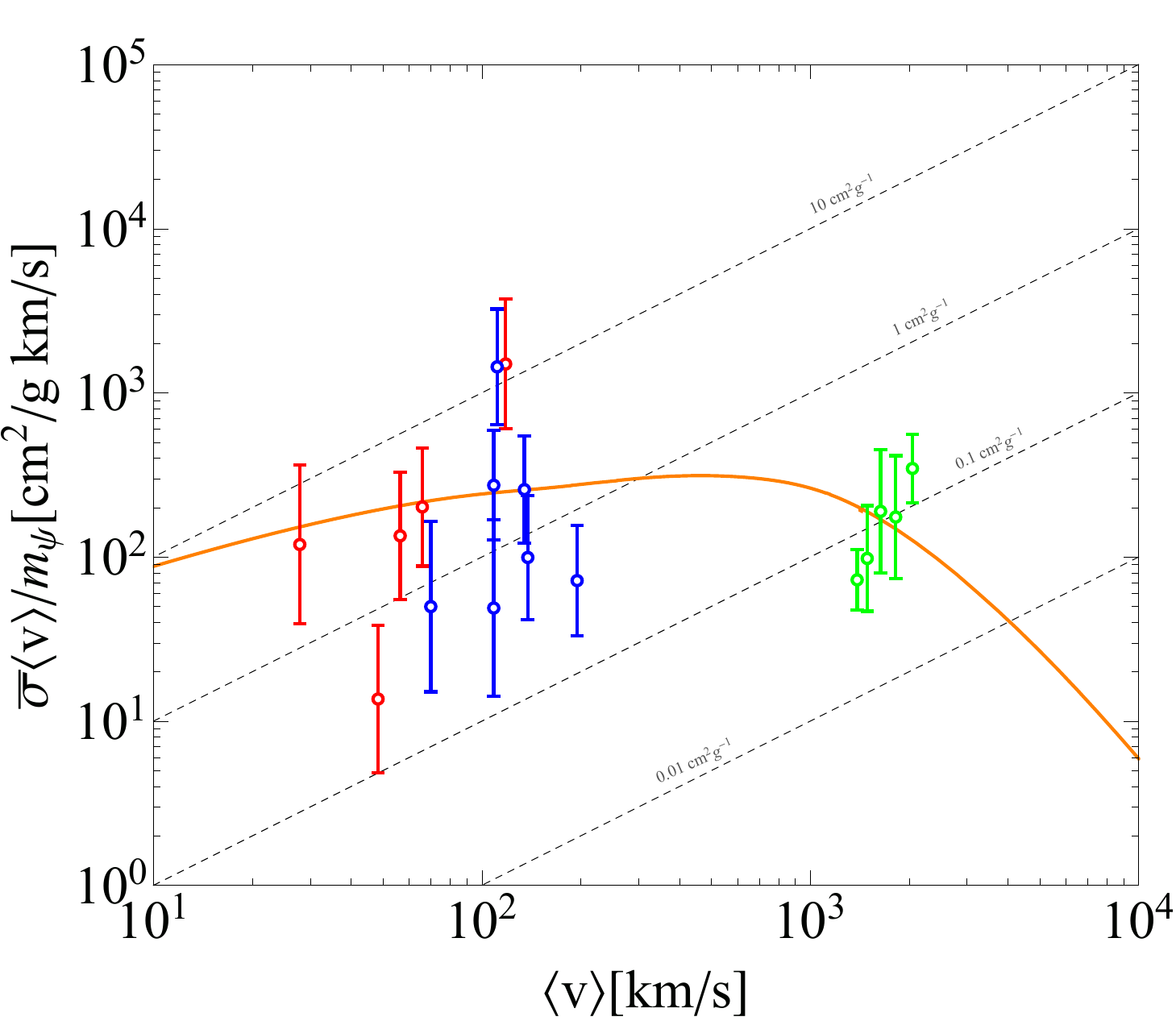}
	\caption{ The averaged cross section of Majorana DM per unit mass $\bar \sigma \langle v_{\rm rel} \rangle / m_\chi$ as a function of $\langle v_{\rm rel}\rangle$ for benchmark point $m_\chi$ = 200 GeV, $m_\phi$ = 5 MeV, $\alpha$ = 0.3. Data~\cite{Kaplinghat:2015aga} inferred from dwarfs (red), LSB galaxies (blue), groups(left green), and clusters (right green) data.
	}
	\label{fig:sigmav}
	\end{figure}

Calculation of velocity averaged viscosity cross-section do not simply follow the procedure of multiplying the $\sigma_V$ by velocity and taking average, instead energy transfer rate related to the viscosity cross-section is defined by~\cite{Colquhoun:2020adl} 

\begin{equation}
    \bar \sigma_V = \frac{\langle \sigma_V v^3_{\rm rel} \rangle}{24/\sqrt{\pi}v_0^3}
\end{equation}
where $v_{\rm rel}$ and $v_0$ are DM relative velocity and velocity dispersion of $v$ which obeys Maxwell-Boltzmann distribution. Averaged cross section per unit mass $\bar \sigma \langle v_{\rm rel} \rangle / m_\chi$ of fermionic DM as a function of $\langle v_{\rm rel}\rangle$ is presented in Fig.~\ref{fig:sigmav} for the benchmark point $m_\chi$ = 200 GeV, $m_\phi$ = 5 MeV, $\alpha$ = 0.3. As is shown in the Fig.~\ref{fig:sigmav}, our model is suitable as a solution to the small scale problems, since the resulting curve is compatible to the data points from five clusters, seven low surface brightness spiral galaxies and six dwarf galaxies. 

\begin{figure}[!htbp]
	\centering
	\includegraphics[width=0.65\textwidth]{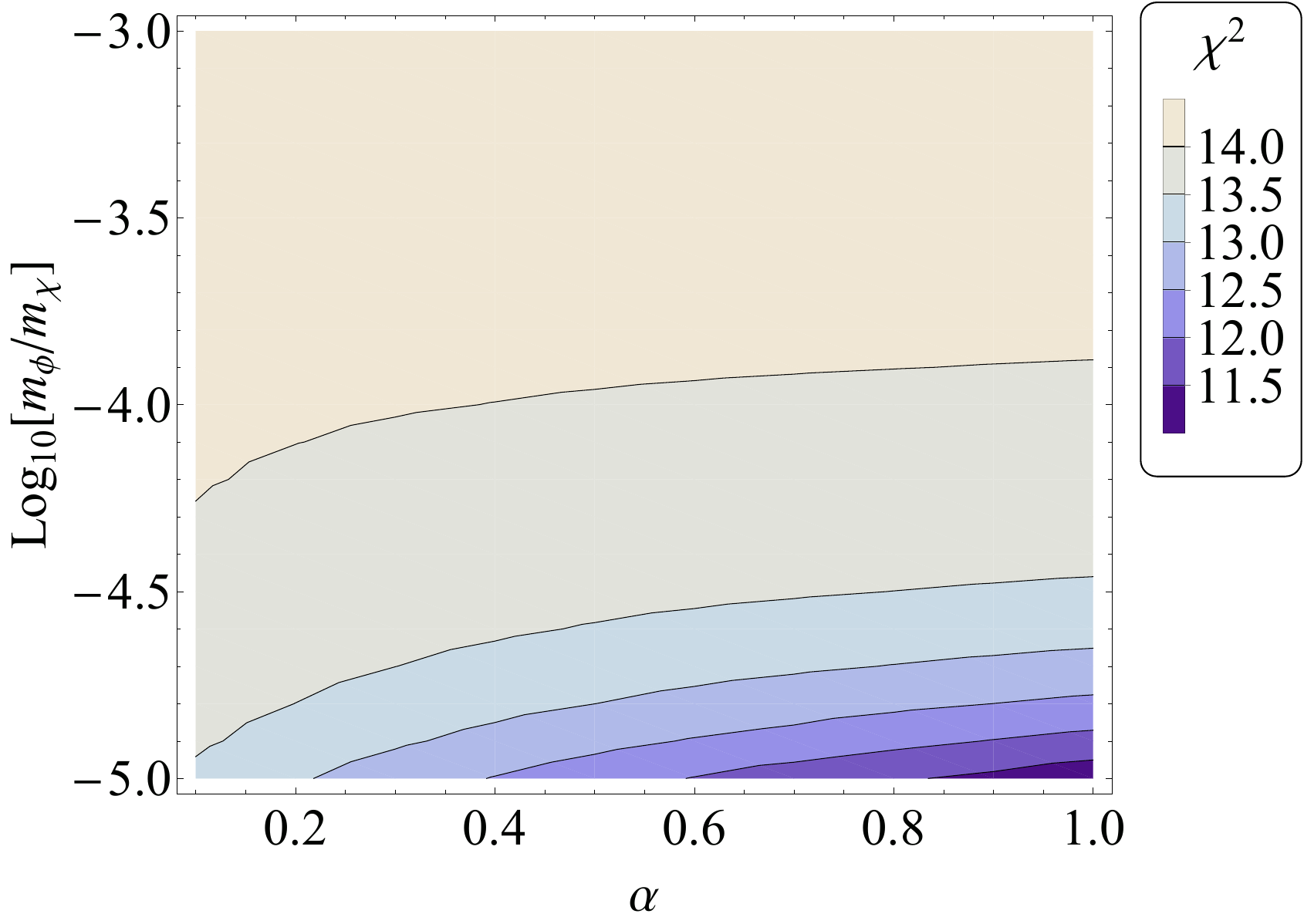}
	\caption{Chi-squared distribution over $\alpha-m_{\phi}/m_{\chi}$ plane. The DM mass is set to be $10^{3}\mathrm{GeV}$.
	}
	\label{fig:chi}
	\end{figure}

There are two relevant parameters in fitting particle input into data $(\eta,\beta)\rightarrow \bar\sigma\langle v\rangle/m_{\chi}$. However $\eta$ and $\beta$ contain the velocity which is not directly relevant to projection on HS. Thus we choose the mass ratio $m_{\phi}/m_{\chi}$ and $\alpha$ as input parameters and perform best fit on this two-parameter model with $m_{\chi}=1$ TeV. Figure~\ref{fig:chi} displays the chi-squared distribution. The value of $\alpha$ has tiny effect on chi-squared which means correct self-interaction cross section is almost independent of $\alpha$ once we fix the value of $m_{\chi}$. The results of the correlation analysis shows that the ratio $m_{\phi}/m_{\chi}$ should be smaller than $10^{-4}$. It is compatible with the semi-classical regime. Throughout the paper we require $m_{\phi}/m_{\chi}<10^{-4}$. 

\section{Cross-Section and Implication on Relic Density}
\label{sec:cross}

In this section we consider $\chi \chi\rightarrow \phi\phi, aa, \phi a$ processes to compute the relic density. It encodes the underlying story on the SE, radiative BSF, and HS freeze-out. Typically, HS freeze-out process includes only born level calculation $\sigma_{\mathrm{eff}}v_{\mathrm{rel}}=\sigma_{\mathrm{tree}}v_{\mathrm{rel}}$. However, non-perturbative corrections from SE and BSF are not negligible when requiring SIDM in Eq.~\eqref{eq:yukawa}. The effective cross-section in our model is thus written as follows

\begin{equation}
\left\langle\sigma_{\text {eff }} v_{\text {rel }}\right\rangle=\left\langle\sigma_{\text {ann }} v_{\text {rel }}\right\rangle+\left\langle\sigma_{\text {BSF }} v_{\text {rel }}\right\rangle_{\text {eff }} ~.
\label{eqn:cross-section}
\end{equation}

The BSF and subsequent decay open up a new effective DM annihilation channel. The simple introduction of singlet offers a promising DM annihilation channel into the mediator. Evaluating the tree-level Feynman diagrams yield the born level cross sections~\cite{Kappl:2010qx} 

\begin{equation}
\begin{aligned}
&\sigma^{\mathrm{ann}}\left(\chi \chi \rightarrow \phi \phi\right)v_{\mathrm{rel}} \simeq \frac{17}{256 \pi} \frac{\kappa^{4}}{m_{\chi}^{2}}\left(1-\frac{22}{51} \frac{A_{\kappa}}{\kappa m_{\chi}}+\frac{1}{17} \frac{A_{\kappa}^{2}}{\kappa^2 m_{\chi}^{2}}\right) v_{\mathrm{rel}}^2 ~, \\
&\sigma^{\mathrm{ann}}\left(\chi\chi \rightarrow a a\right)v_{\mathrm{rel}} \simeq \frac{9}{256 \pi} \frac{\kappa^{4}}{m_{\chi}^{2}}\left(1-\frac{14}{27} \frac{A_{\kappa}}{\kappa m_{\chi}}+\frac{1}{9} \frac{A_{\kappa}^{2}}{\kappa^2 m_{\chi}^{2}}\right)v_{\mathrm{rel}}^2 ~, \\
&\sigma^{\mathrm{ann}}\left(\chi\chi \rightarrow \phi a\right)v_{\mathrm{rel}} \simeq \frac{9}{64 \pi} \frac{\kappa^{4}}{m_{\chi}^{2}}\left(1+\frac{2}{3} \frac{A_{\kappa}}{\kappa m_{\chi}}+\frac{1}{9} \frac{A_{\kappa}^{2}}{\kappa^2 m_{\chi}^{2}}\right) ~.
\end{aligned}
\end{equation}

The first two processes are p-wave contribution  $\sigma_p v_{\mathrm{rel}}$ and the third one is s-wave cross-section $\sigma_s v_{\mathrm{rel}}$. The expression is valid only in large $m_{\chi}/m_{\phi,a}$ limit. The long-range effect is re-summed to solve the Schrodinger equation with Yukawa potential $V(r)$. The short distance annihilation cross-section is encoded in the absorptive part of forward scattering amplitude called Wilson coefficient $f[{}^{2s+1}L_j]$. The schematic form of s-wave Sommerfeld enhanced cross-section is written as $\sigma_{\mathrm{eff}}v_{\mathrm{rel}}=\sigma_{\mathrm{tree}}v_{\mathrm{rel}}S_0$. In the limit of Hulthen potential, we can present SE in s-wave and p-wave by

\begin{equation}
S_0(v_{\mathrm{rel}} )=\frac{\frac{\pi}{\varepsilon_{v}} \sinh \left[\frac{12 \epsilon_{v}}{\pi \varepsilon_{\phi}}\right]}{\cosh \left[\frac{12 \varepsilon_{v}}{\pi \varepsilon_{\phi}}\right]-\cos \left[2 \pi \sqrt{\left.\frac{6}{\pi^{2} \varepsilon_{\phi}}-\left(\frac{6 \varepsilon_{v}}{\pi^{2} \varepsilon_{\phi}}\right)^{2}\right]}\right.}
\end{equation}
and 

\begin{equation}
S_{1}(v_{\mathrm{rel}} )=\frac{\left(1-\varepsilon_{\phi} \pi^{2} / 6\right)^{2}+4 \varepsilon_{v}^{2}}{\left(\varepsilon_{\phi} \pi^{2} / 6\right)^{2}+4 \varepsilon_{v}^{2}} S_{0} 
\end{equation}
respectively. Here $\varepsilon_{v} \equiv v_{\mathrm{rel}} /\left(2 \alpha\right)$ and $\varepsilon_{\phi} \equiv m_{\phi} /\left(\alpha m_{\chi}\right)$. As a result, the thermal averaged cross-section with the SE is obtained by

\begin{equation}
\left\langle\sigma_{\mathrm{ann}} v_{\mathrm{rel}}\right\rangle=\frac{x^{3 / 2}}{2 \sqrt{\pi}} \int_{0}^{\infty} d v_{\mathrm{rel}} v_{\mathrm{rel}}^{2}\left(\sigma_{s} v_{\mathrm{rel}} S_0(v_{\mathrm{rel}})+\sigma_p v_{\mathrm{rel}} S_1(v_{\mathrm{rel}} )\right) e^{-x v_{\mathrm{rel}}^{2} / 4} ~.
\end{equation}

In most parameter spaces, the contributions of the s-wave predominate over the p-wave, although it suffers a dangerous constraint from CMB ionization. However, we propose a blind spot scenario  $A_{\kappa}=-3\kappa m_{\chi}$, where the s-wave cross-section vanishes automatically. The CMB constraint is thus alleviated.  However, the BSF cross-section remains s-wave to re-introduce CMB constraint, even though it is small.

The cross section of radiative BSF in ground state, i.e. $n=1,l=m=0$ is~\cite{Oncala:2018bvl}
\begin{equation}
    \sigma_{\mathrm{BSF}}v_{\mathrm{rel}}=2\frac{\alpha^3}{\mu^2}\left(\frac{2A_{\kappa}}{\mu \alpha^2}\right)^2\sqrt{1-\left(\frac{2m_{\phi}}{(\alpha^2+v_{\mathrm{rel}^2})}\right)^2}S_{0}^{\mathrm{BSF}}(\zeta,\xi)\left(\frac{\zeta^2}{1+\zeta^2}\right)^3\exp(-4\zeta\arccot(\zeta)) ,
    \label{eqn:BSF}
\end{equation}
where the characteristic parameter $\zeta=\alpha/v_{\mathrm{rel}}$ represents the ratio between Bohr momentum and momentum exchange of unbound particles, while $\xi=\mu\alpha/0.84m_{\phi}$ is the ratio of interaction range and Bohr radius. Both are larger than $1$ to allow for the existence of the bound state, existing as phase space suppression in the following formula. The effectively BSF cross-section is also the function of the ionization and decay process
\begin{equation}
\left\langle\sigma_{\mathrm{BSF}} v_{\mathrm{rel}}\right\rangle_{\mathrm{eff}}=\left\langle\sigma_{\mathrm{BSF}} v_{\mathrm{rel}}\right\rangle \times\left(\frac{\Gamma_{\mathrm{dec}}}{\Gamma_{\mathrm{dec}}+\Gamma_{\mathrm{ion}}}\right) ~,
\end{equation}
where ionization rate of bound states is
\begin{equation}
\Gamma_{\mathrm{ion}} =\left\langle\sigma_{\mathrm{BSF}} v_{\mathrm{rel}}\right\rangle\left(\frac{m_{\chi} T_h}{4 \pi}\right)^{3 / 2} e^{-\left|E_{\mathrm{B}}\right| / T_h} ~.
\end{equation}
The binding energy $E_\mathrm{B}$ determines the relative strength of the ionization rate. The annihilation decay is responsible for the decay of the bound state where we replace the Sommerfeld factor by the squared of bound state wave-function
\begin{equation}
\Gamma_{\mathrm{dec}} =\left|\Psi_{100}(0)\right|^{2}\left(\sigma_{\mathrm{ann}} v_{\mathrm{rel}}\right)^{\mathrm{tree}}.
\label{eqn:decay}
\end{equation}
The Eq.~\eqref{eqn:decay} is only valid at the s-wave dominated case, since the conservation of angular momentum forbids the p-wave decay process. Instead, the only allowed decay channel is $4\phi$ process $\mathcal{B}\rightarrow\phi\phi\phi\phi$,
\begin{equation}
\Gamma_{\mathcal{B} \rightarrow 4 \phi}=\frac{0.01\left|\Psi_{100}(0)\right|^{2} \kappa^8}{49152 \pi^{6} m_{\chi}^{2}} ~.
\end{equation}

\begin{figure}[!htbp]
	\centering
	\includegraphics[width=0.45\textwidth]{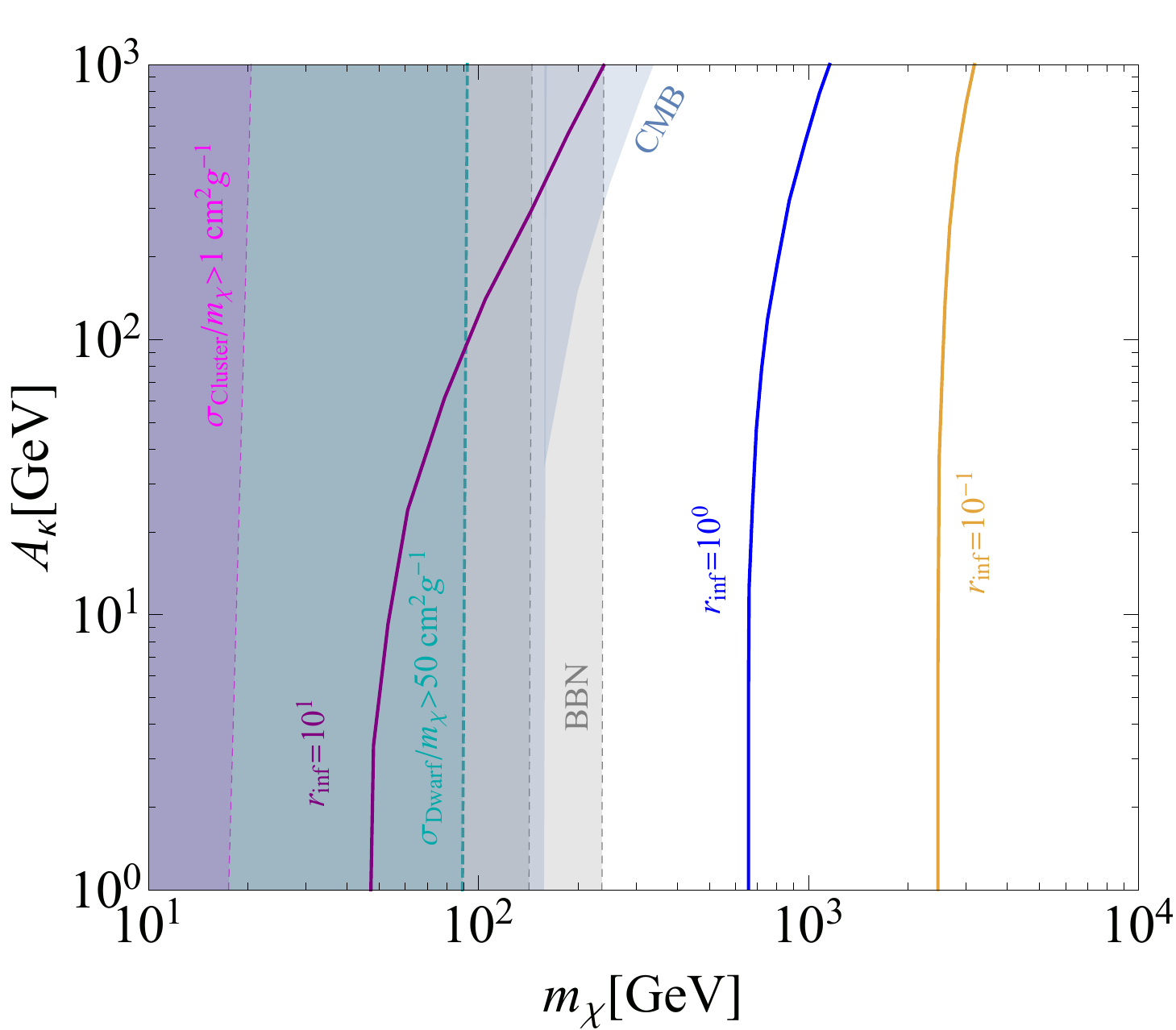} 
	\includegraphics[width=0.45\textwidth]{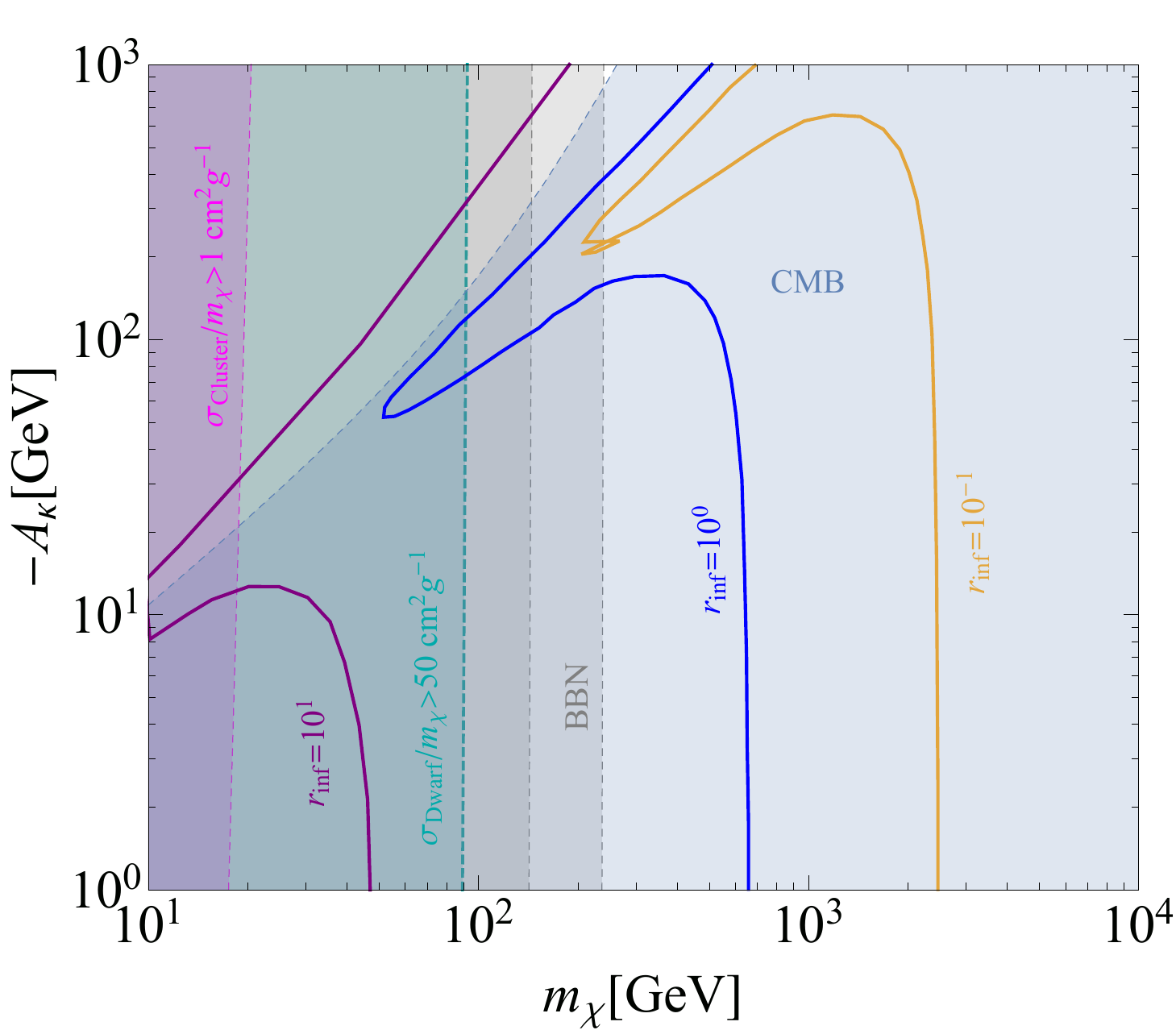}
		\includegraphics[width=0.45\textwidth]{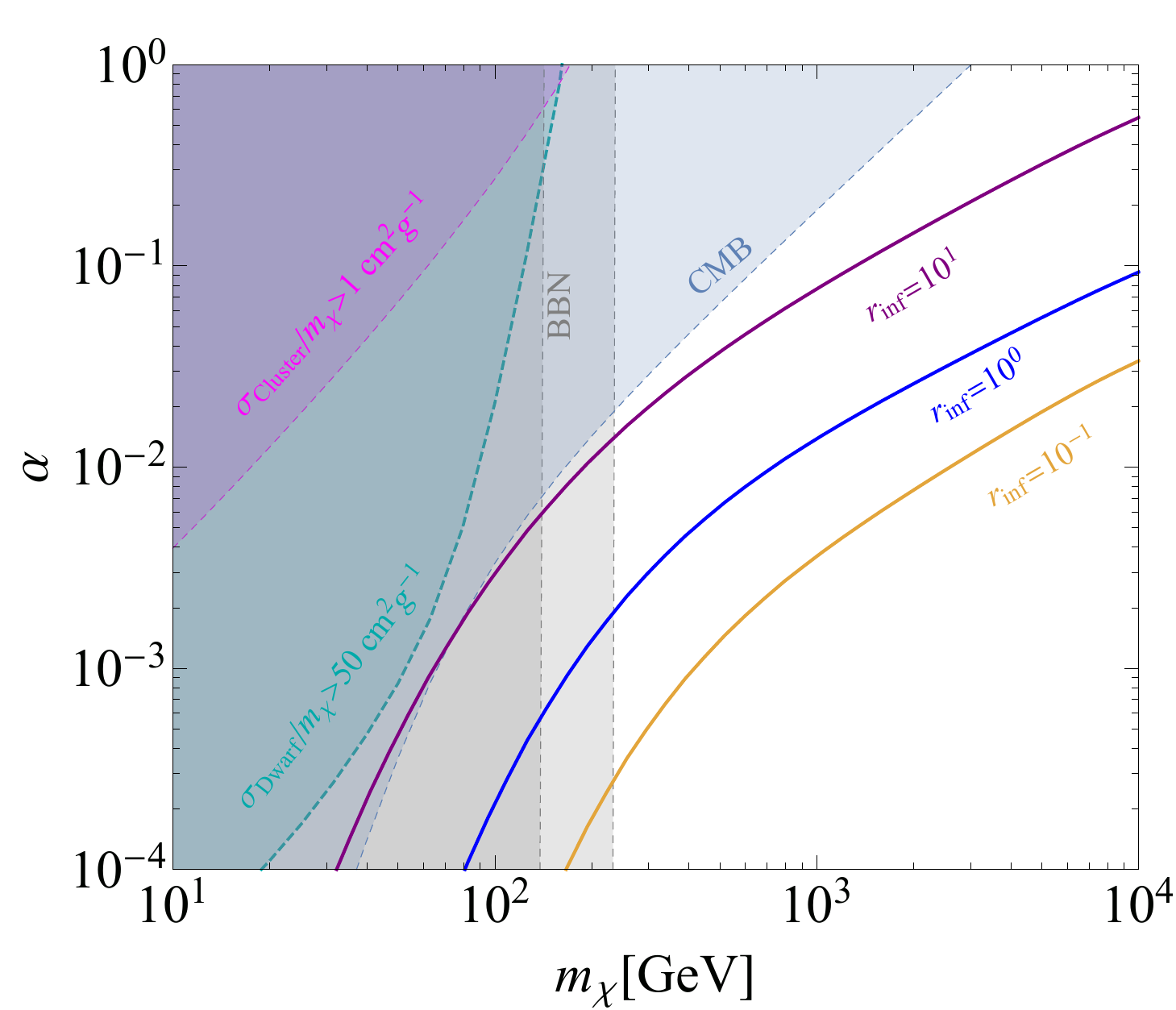}
	\caption{Constraints on the parameter space. The solid lines satisfy relic density of $\Omega h^2 = 0.11$ for different $r_\mathrm{inf}$. The Green and magenta-shaded regions are constrained from dwarf and cluster scale SIDM cross-sections. The gray shaded regions indicates where mediator decays occur after neutrino decoupling and is excluded by BBN. The light blue shaded regions are ruled out by CMB constraints. The $\alpha$ and $A_\kappa$ are fixed to $0.01$ and 100 GeV on the top and bottom panels, respectively. The mass ratio is $m_{\phi}/m_{\chi}=10^{-5}$.}
	\label{fig:FigRelic}
\end{figure}

The Boltzmann equation in Eq.~\eqref{eqn:Boltzmann} can be numerically solved for the cross section in Eq.~\eqref{eqn:cross-section} to obtain DM relic density. In Fig.~\ref{fig:FigRelic}, we give the solid lines which satisfy correct relic density, constraints from various experiments are also presented. We can obtain the correct relic density in blue lines where the initial condition on reheating is $r_{\mathrm{inf}}=1$. To show the deviation of parameter space, we also consider different values of $r_{\mathrm{inf}}=10^{-1},10^{1}$ with purple and orange lines. In the bottom panel, we find larger $r_{\mathrm{inf}}$ requires larger interaction strength $\alpha$ and is ruled out by CMB constraint. Large $r_{\mathrm{inf}}$ induces a large effective Hubble parameter to obtain correct relic density, a larger annihilation cross-section is required that favors larger $\alpha$. The panel on the top left shows that positive $A_{\kappa}$ does not affect relic density to a large extent, since cross-section is almost the function of $\alpha$ and $m_{\chi}$. For negative $A_{\kappa}$ on the top right panel, the cancellation between $A_{\kappa}$ and $m_{\chi}$ will make the cross-section smaller than the required one. As a result, light DM compensates for the cancellation effect. When negative $A_{\kappa}$ becomes smaller than $m_{\chi}$, it will be the same as positive $A_{\kappa}$. In addition, the CMB constraint rules out the $r_{\mathrm{inf}}$ larger than $10$ for the positive $A_{\kappa}$. While negative $A_{\kappa}$ scenario almost does not have a viable parameter space after implementing the CMB and the BBN constraints. 
The observation data from the SIDM, BBN and CMB set strong constraints on the parameter space. We demonstrate these constraints by showing the shaded excluded regions in the Fig~\ref{fig:FigRelic}: 
\begin{itemize}
    \item Green-shaded region for dwarf. \\
    To solve small-scale issues, a large self-interaction cross-section in dwarf galaxies is required. However, it can not be arbitrarily large. Instead astrophysical data puts an upper bound on self-interaction cross-section on dwarf scale i.e. $\sigma/m_{\chi}<50~\mathrm{cm}^2\mathrm{g}^{-1}$. The bottom panel displays the weak dependence of self-interaction cross-section on $\alpha$ and $m_{\chi}$. While $A_{\kappa}$ plays no role in SIDM, the vertical line in the top panel indicates the sensitivity of SIDM on $m_{\chi}$.
    
    \item Magenta-shaded region for cluster.\\
    On the cluster scale, according to the astrophysical observations, the self-interacting cross-section must reduce to be smaller than $1~\mathrm{cm}^2\mathrm{g}^{-1}$.
    
    \item Gray shaded region for BBN.\\
   The SIDM works well in the HS itself, which is not sensitive to different temperature evolution. While BBN is the function of decay width of mediator, it is sensitive to the $r_{\mathrm{inf}}$. The decoupling from kinetic equilibrium can be realized by selecting $\epsilon=10^{-8}$ for $r_{\mathrm{inf}}=10^{-1},10^0$ and $\epsilon=6\times 10^{-9}$ for $r_{\mathrm{inf}}=10$. Since we choose the ratio $m_{\phi}/m_{\chi}=10^{-5}$ in both figures, the constraint on a mediator from BBN can be interpreted as the constraint on DM mass. Therefore we have two vertical lines for the BBN boundary. It is easy to find that the BBN constraint is the most strict one for light DM. 
    
    \item Blue-shaded region for CMB\\
  The DM annihilation products inject energy into the CMB and lead to anisotropies. It provides a sensitive probe of DM annihilation during the dark ages. The s-wave cross-section is bounded from above \cite{Slatyer:2015jla}, in order not to distort the CMB spectrum, which is roughly~\cite{An:2016kie}
    \begin{equation}
\lim _{v \rightarrow 0}(\sigma v)<3 \times 10^{-24} \mathrm{~cm}^{3} \mathrm{sec}^{-1} \times\left(\frac{m_{\chi}}{\mathrm{TeV}}\right) .
\label{eqn:CMB}
\end{equation}

\end{itemize}

\begin{figure}[!htbp]
	\centering
	\includegraphics[width=0.65\textwidth]{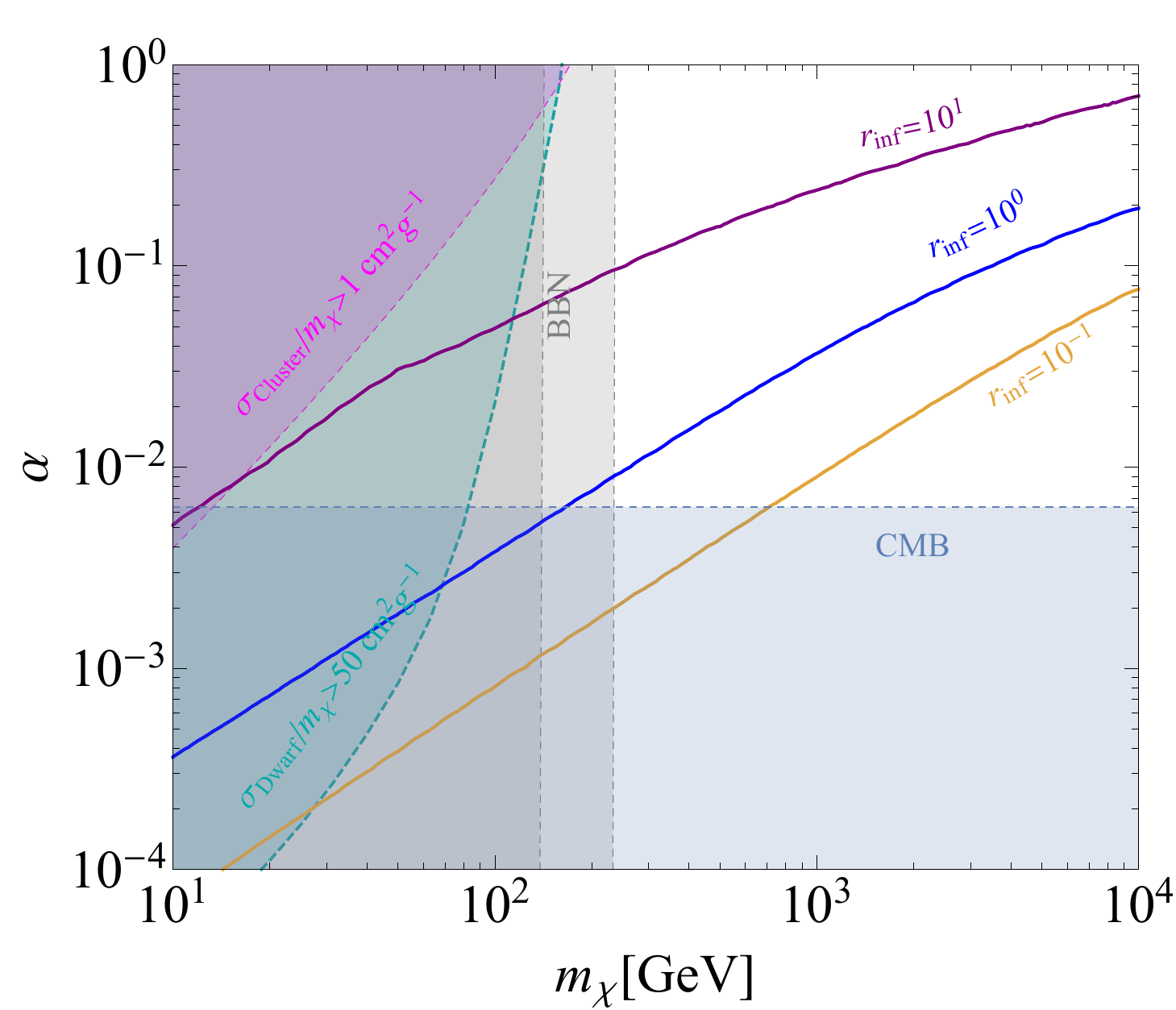}
	\caption{Same as Fig. \ref{fig:FigRelic}, but for the blind spot, i.e. $A_\kappa = -3\kappa m_\chi$.}
	\label{fig:FigRelic2}
\end{figure}

Even though the negative $A_{\kappa}$ leads to null parameter space for our model, there is still a blind spot, in which the s-wave annihilation automatically vanishes and the p-wave annihilation process dominates. Therefore, the CMB constraint becomes relevant only if the BSF cross-section becomes large.  The task, for now, is to compute the limit of $v\rightarrow 0$ for BSF cross-section. The SE in Eq.~\eqref{eqn:BSF} is 

\begin{equation}
S_{\mathrm{BSF}}^0=\left|\frac{\Gamma\left(a^{-}\right) \Gamma\left(a^{+}\right)}{\Gamma(1+2 i w)}\right|^{2} ~,
\end{equation}
where $a^{\pm}=1+i w(1 \pm \sqrt{1-x /\omega})$, $\omega=m_{\chi}v_{\mathrm{rel}}/2m_{\phi}$ and $x=2\alpha/v_{\mathrm{rel}}$. When the velocity is smaller than $m_{\phi}/m_{\chi}$, it has a non-vanishing limit

\begin{equation}
\begin{aligned}
(S_{\mathrm{BSF}}^0)_{v_{\mathrm{rel}}\rightarrow 0}&=\Gamma\left[1-\sqrt{\frac{{m}_{\chi} \alpha}{{m}_{\phi}}}\right]^{2} \Gamma\left[1+\sqrt{\frac{{m}_{\chi} \alpha}{{m}_{\phi}}}\right]^{2}\\
&=\left(\Gamma\left[1-\sqrt{\frac{{m}_{\chi} \alpha}{{m}_{\phi}}}\right] \times \Gamma\left[\sqrt{\frac{{m}_{\chi} \alpha}{{m}_{\phi}}}\right] \sqrt{\frac{{m}_{\chi} \alpha}{{m}_{\phi}}}\right)^{2} \\
&=\left(\frac{\pi}{\sin \left[\pi \sqrt{\frac{{m}_{x} \alpha}{{m}_{\phi}}}\right]} \sqrt{\frac{{m}_{\chi} \alpha}{{m}_{\phi}}}\right)^{2} ~.
\end{aligned}
\end{equation}
By substituting into the s-wave BSF cross-section, we obtain the analytical cross-section

\begin{equation}
    (\sigma_{\mathrm{BSF}})_{v_{\mathrm{rel}}\rightarrow 0}=\frac{4608\alpha\pi^3 \sqrt{1-(16m_{\phi}^2/\alpha^4 m_{\chi}^2)}}{e^4 m_{\phi}m_{\chi}\sin^2(\pi\sqrt{\alpha m_{\chi}/m_{\phi}})} ~.
\end{equation}
Generally, it is larger than the CMB bound in Eq.~\eqref{eqn:CMB}. The loophole is that the BSF cross-section vanishes when the phase space is not open, i.e. $\alpha<6.32\times 10^{-3}$, which we use it to constrain the coupling in Fig.~\ref{fig:FigRelic2}. 

Finally, we show the benchmark points that satisfy both relic density and constraints in Tab.~\ref{tab:benchmarks}. Even though, large cancellation between negative $A_{\kappa}$ and $m_{\chi}$ can increase the fraction of BSF cross-section, BBN prevents it to be larger than $5\%$.
\begin{table}[htbp]
\begin{centering}
\begin{tabular}{|c|ccccc|} \hline  
Benchmarks     & 
$m_{\chi}$ (GeV)        &  
$A_{\kappa}$ (GeV)    & 
$\alpha$         & 
$\sigma_{\mathrm{ann}}$ ($\mathrm{GeV}^{-2}$)  & 
$f$   \\ \hline   
BP1   & 
$0.5\times 10^3$   & 
$10^2$  & 
$6\times 10^{-3}$   & 
$1.57\times 10^{-9}$   & 
$1\%$\\ \hline
BP2  & 
$0.52\times 10^3$    &
$-1.2\times 10^{2}$  & 
$10^{-2}$     & 
$1.63\times 10^{-9}$   & 
$4\%$
\\ \hline
\end{tabular}
\caption{Benchmarks for realizing correct relic density without any constraints. We set $r_{\mathrm{inf}}=1$ for simplicity. The $f$ is fraction of BSF cross-section over total cross-section i.e. $f=\langle\sigma_{\mathrm{BSF}}v_{\mathrm{rel}}\rangle/\langle\sigma_{\mathrm{total}}v_{\mathrm{rel}}\rangle$.}
\label{tab:benchmarks} 
\end{centering}
\end{table}

It is necessary to mention that, from observational perspective, long-lived mediator in the hidden sector still has a gamma-ray indirect detection signature~\cite{1709.07002, 2106.09740}. The DM capture~\cite{1703.04629} and annihilation in the Sun also can produce detectable gamma-ray for a long-lived mediator. Unfortunately, these observational consequences would not, as many other DM models, differentiate our model from others which predict same signal detection.

\section{Conclusion}
\label{sec:conclusion}

The purpose of the paper is to explore the long-range potential effect in the hidden sector. It provides the correct self-interaction cross-section and a new s-wave annihilation contribution from the bound state formation. Our study quantifies the influence of different temperature ratios $r_{\mathrm{inf}}$ on relic density in General Higgs portal dark matter model. Even though the bound state formation cross-section is not sufficient to affect relic density, the existence of the new s-wave contribution plays a crucial role in CMB constraint. For the positive $A_{\kappa}$, the combination of CMB and BBN constraint favors the heavy dark matter and not-too-hot hidden sector. For the negative $A_{\kappa}$, no parameter space for the model except for the blind spot scenario that we propose.

\section*{Acknowledgements}
We would like to thank the anonymous referees for the very helpful suggestions that improved this study. This work is supported by the National Natural Science Foundation of China (11805161, 12047560), by the Basic Science Research Program through the National Research Foundation of Korea (NRF) funded by the Ministry of Education, Science and Technology (NRF- 2019R1A2C2003738), and by the Korea Research Fellow-ship Program through the NRF funded by the Ministry of Science and ICT (2019H1D3A1A01070937), and by China Postdoctoral Science Foundation (2020M681757).

\bibliographystyle{JHEP}
\bibliography{lit}

\end{document}